\documentclass[reprint,
 amsmath, amssymb,
 aps, floatfix, showkeys
]{revtex4-2}

\usepackage{graphicx}
\usepackage{dcolumn}
\usepackage{bm}
\usepackage[colorlinks, linkcolor = magenta, urlcolor = blue, citecolor = blue]{hyperref}
\usepackage{braket}
\usepackage{physics}
\usepackage{amsmath}
\usepackage{array}
\usepackage{float}
\usepackage{multirow}

\begin{document}

\title{Non-classicality of two-qubit quantum collision model: non-Markovian effects}
\author{Jai Lalita}
\email{jai.1@iitj.ac.in}
\author{Subhashish Banerjee}
\email{subhashish@iitj.ac.in }
\affiliation{Indian Institute of Technology, Jodhpur-342030, India}

\date{\today}

\begin{abstract}
We investigate a two-qubit quantum system in contact with an environment modeled by a microscopic collision model with added ancilla-ancilla collisions in the non-Markovian regime. Two schemes of the two-qubit collision model with carried-forward correlations are introduced. In one scheme, a single stream of ancillae interacts with only one of the qubits of the two-qubit system; in the other, both the qubits interact with two independent sequences of ancillae, which could be at the same or different temperatures. The system's non-Markovian evolution is examined using the trace distance measure, and the non-classicality of the system is studied using the Wigner function, non-classical volume, and concurrence. Also, interesting steady-state behavior is observed when both the independent ancillae are kept at the same temperature.
\end{abstract}

\keywords{Quantum collision model, open quantum systems, non-Markovianity, non-classicality}

\maketitle
\section{\label{sec:intro}Introduction\protect}
The theory of open quantum systems seeks to characterize and comprehend the dynamical behavior of quantum systems that are inherently coupled to their surroundings~\cite{Quantum_Dissipative_Systems, Breuer2007, Banerjee2018, Vacchini_2011, tiwari2024strong}. Such interactions between a system and its environment are unavoidable and result in a nonunitary evolution of the system's state. This interaction typically causes decoherence and dissipation of quantum information from the system of interest. A widely adopted formalism for analyzing open quantum systems is the Gorini–Kossakowski–Sudarshan–Lindblad (GKSL) master equation, which provides a mathematical description of the system's Markovian dynamics under the Born–Markov and rotating wave approximations~\cite{GKLSpaper, Lindblad1976}. Theoretical developments and experimental advances have enabled the exploration of more general non-Markovian dynamics in open quantum systems, attracting considerable attention in recent years~\cite{Hall_2014, Rivas_2014, RevModPhys.88.021002, CHRUSCINSKI20221, banerjeepetrucione, vega_alonso, Shrikant2018non_Markovianity, Utagi2020}.

However, the mathematical framework for describing open quantum systems can become cumbersome, both in the derivation and solution of the equations regulating the system's dynamics. An alternate method that is easy to understand, adaptable, and simple for tracking the time evolution of the open system is the collision model approach~\cite{CM_1963, Ciccarello2022Quantumcollisionmodel}. Collision models have been extensively used in recent years to tackle various problems like decoherence, dissipation, and non-Markovian behavior in a controlled and tractable way~\cite{ziman2005description, Rybár_2012, Ciccarello2013Collision-model, McCloskey2014Non-Markovianity, CM_nm_2016, Kretschmer2016Collisionmodel, CM_nm_2017, Campbell2018Systemenvironment, Rodrigues2019thermodynamics, Landi2021Irreversibleentropy, csenyacsa2022entropy}. They have also found applications in quantum thermodynamics for modeling thermalization and heat exchange processes~\cite{Strasberg2017Quantum, Pezzutto2016Implications, Campbell2021Collision, DeChiara_2018Reconciliation}.

Moreover, the two-qubit collision model provides a more comprehensive framework than its single-qubit counterpart for simulating open quantum dynamics, as it captures both inter-qubit correlations and system-environment interactions. Unlike single-qubit models, which are restricted to individual decoherence processes, the two-qubit setting enables the study of entanglement dynamics, collective behavior, and environment-induced correlations between system qubits, which are key features in quantum communication and computation protocols. These advantages make the two-qubit collision model particularly suitable for exploring realistic quantum systems where multi-qubit systems' coherence and interaction with the environment are central. 

In this study, we examine such a two-qubit collision model by simulating a sequence of ``collisions" between the two-qubit system and environmental ancillae to describe the system-environment interaction in a controllable manner. Depending on how the environmental degrees of freedom interact, such a ``collision" model can simulate both Markovian and non-Markovian dynamics~\cite{ThermalizingQuantumMachines_2002, ziman2005description, Rybár_2012, McCloskey2014Non-Markovianity, Ciccarello2013Collision-model, Kretschmer2016Collisionmodel}. We also introduce two different schemes to investigate the effects of manipulating the interactions between environmental ancillae and the two-qubit system on the non-Markovianity and non-classicality of the two-qubit system. For that, we study trace distance, Wigner function, non-classical volume, and dynamics of entanglement generation of the two-qubit system using both schemes. The trace distance is diagnostic for non-Markovianity witnessed by information flow, while the Wigner function and non-classical volume reveal phase-space signatures of quantum behavior and deviations from classicality~\cite{Wigner1932Quantum, zavatta2004quantum, Thapliyal2015Quasiprobability}. Moreover, entanglement generation with the number of collisions allows the identification of regimes where quantum correlations are enhanced or degraded by the environment~\cite{Naikoo2019Facets, Tiwari2023QuantumCorrelations}. We also investigate the steady state behavior under varying physical conditions for possible thermodynamic applications~\cite{Hanggi_talkner_review, tiwari2024strong, Thomas2018thermodynamics, Ashutosh2023thermodynamics}.

The article is organized as follows. In Sec.~\ref{Collision model}, we introduce the two-qubit collision model and two different interaction schemes, \textit{viz}. scheme A and scheme B. Section~\ref{Non_Markovianity_measure} investigates the non-Markovian nature of the model through a non-Markovianity witness, i.e., trace distance, for both the schemes. It is followed by examining both schemes' non-classical features in Sec.~\ref{Wigner_function} and~\ref{Quantum_correlation}. In Sec.~\ref {Wigner_function}, we analyze the Wigner function and the associated non-classical volume. Section~\ref {Quantum_correlation} explores the emergence of quantum correlations, particularly quantum entanglement, using both schemes for initially separable two-qubit states. Additionally, the steady-state behavior of the system qubits is analyzed in Sec.~\ref{SS_behaviour} for scheme B. Finally, our conclusions are summarized in Sec.~\ref{conclusion}.

\section{\label{Collision model} Collision model}
The single-qubit collision model is a powerful framework for studying how a quantum system interacts with its environment. In this model, the system qubit interacts one at a time with a sequence of identical environmental qubits (often called ancillae) through unitary operations~\cite{Ciccarello2022Quantumcollisionmodel, McCloskey2014Non-Markovianity, Ciccarello2013Collision-model, Campbell2018Systemenvironment, Campbell2021Collision, csenyacsa2022entropy}. When the system interacts with individual environmental qubits and then moves on to interact with fresh ones while discarding the previous ones, its behavior follows a Markovian process. However, memory effects can arise if an interaction is introduced between consecutive environmental qubits, especially between the one that just interacted with the system and the next one. This allows some of the system's lost information to be partially retrieved later in the process. These memory effects play a key role in the system's dynamics, which we will explore further. The system's evolution over time is determined by repeatedly following these interaction steps while tracking its state. The flexibility of collision models is particularly valuable, as they allow adjustments to the number of qubits and the nature of interactions, making them adaptable to different problems~\cite{Cattaneo2021Collision, Cattaneo2022ABriefJourney}. 

Here, we consider a collision model for a two-qubit system $S$, composed of qubits $s_1$ and $s_2$, interacting with a stream of identical ancillae one at a time. The system-ancilla interaction is modeled using two schemes,~\textit{viz.} scheme A and scheme B, as depicted in Figs.~\ref{approach_1} and \ref{Scheme_B}, respectively. In scheme A, a stream of ancillae denoted as $a_n^{R}$ interacts with one of the system qubits say $s_2$, whereas in scheme B two independent streams of ancillae denoted by $a_n^{L}$ and $a_n^{R}$ interact with the qubits $s_1$, and $s_2$ of the system, respectively. Throughout the paper, the ancillae are considered to be initially in the thermal state 
\begin{eqnarray}\label{eq_ancilla_thermal_state}
    \rho_{0}^{a_n^{e}} = e^{-\beta_{a_n^{e}}H_{a_n^{e}}}/\Tr[e^{-\beta_{a_n^{e}}H_{a_n^{e}}}], ~~~~~\forall ~e \in \{L, R\},
\end{eqnarray}
where $(L,~R)$ denotes the left ($L$) and right ($R$) sides of the two-qubit system, and $\beta_{a_n^{e}} = 1/k_BT_{a_n^{e}}$ are the inverse temperatures of the environment qubits, and we set $\hbar = k_B = 1$.

The Hamiltonians that describe the system qubits and environment ancillae are defined as
\begin{eqnarray}
\nonumber
    H_{s_1} = \hbar\omega_{s_1}\sigma_z,
    ~H_{s_2} = \hbar\omega_{s_2}\sigma_z,\\
    H_{a_n^{L}} = \hbar\omega_{a_n^{L}}\sigma_z,
    ~H_{a_n^{R}} = \hbar\omega_{a_n^{R}}\sigma_z.
\end{eqnarray}
Here, $\omega_{s_1}$, $\omega_{s_2}$, $\omega_{a_n^{L}}$, and $\omega_{a_n^{R}}$ are the corresponding transition frequencies of the system and environment qubits, and $\sigma_z$ is the Pauli $Z$ matrix. Throughout the paper, except stated otherwise, we assume that the system qubits and ancillae are in resonance, $\it{i.e.}$, $\omega_{s_1} = \omega_{s_2} = \omega_{a_n^{L}} = \omega_{a_n^{R}} = 1$. The intra-system and system-ancilla interactions of the two-qubit system are governed by the Heisenberg interaction as follows
\begin{eqnarray}
    H_{s_1s_2} &=& g_{s_1s_2}(\sigma_x^{s_1}\sigma_x^{s_2} + \sigma_y^{s_1}\sigma_y^{s_2}),\nonumber \\
    H_{s_ja_n^{e}} &=& g_{s_ja_n^{e}}(\sigma_x^{s_j}\sigma_x^{a_n^{e}} + \sigma_y^{s_j}\sigma_y^{a_n^{e}}), ~~\forall~ j \in \{1, 2\},
    \label{H_int_unitary}
\end{eqnarray}
where $\sigma_{x,y}$ are the Pauli matrices and $g_{s_1s_2}$, $g_{s_ja_n^{e}}$ are the intra-system and system-ancilla coupling constants, respectively, and $e$ can be $L$ or $R$. Additionally, the Hamiltonian for the swap operation that governs the intra-ancilla interactions is
\begin{equation}
    H_{SWAP} = \frac{1}{2}(\Vec{\sigma}{^{a_n^{e}}}.~\Vec{\sigma}{^{a_{n+1}^{e}}} + I_4),
\end{equation}
where $\Vec{\sigma}{^{a_n^{e}}}$ and $\Vec{\sigma}{^{a_{n+1}^{e}}}$ are the Pauli matrices of the two consecutive ancillae that are swapped and $I_4$ is the $4 \cross 4$ identity operator. The corresponding unitary operator is
\begin{equation}
    \begin{aligned}
        U_{a_n^{e}, a_{n+1}^{e}} = e^{-i \Theta H_{SWAP}} = \cos(\Theta) I - i\sin(\Theta)H_{SWAP},   
    \end{aligned}
    \label{p_swap_unitary}
\end{equation}
where $e \in \{L,~R\}$, and $\Theta \in [0, \frac{\pi}{2}]$ is the intra-ancilla interaction strength that can be used to alter from Markovian to non-Markovian behavior. In other words, it can introduce and regulate memory effects~\cite{McCloskey2014Non-Markovianity, ThermalizingQuantumMachines_2002}. 

Further, the time evolution of the two-qubit system is governed by the consecutive application of three unitaries, which include the intra-system, system-ancilla, and intra-ancilla interactions, where the latter's presence is necessary for introducing non-Markovianity into the dynamics~\cite{CM_nm_2016, CM_nm_2017, Campbell2018Systemenvironment}. In both schemes, we keep the evolved joint state of the system and ancilla $\sigma^{Sa^e}_{n + 1}$, see for example Eq.~\eqref{joint_state} below, untouched and use it, as it is, in the next iteration, summarized in detail below.  
\subsection{Scheme A}
\begin{figure}
    \centering
    \includegraphics[width=1\linewidth]{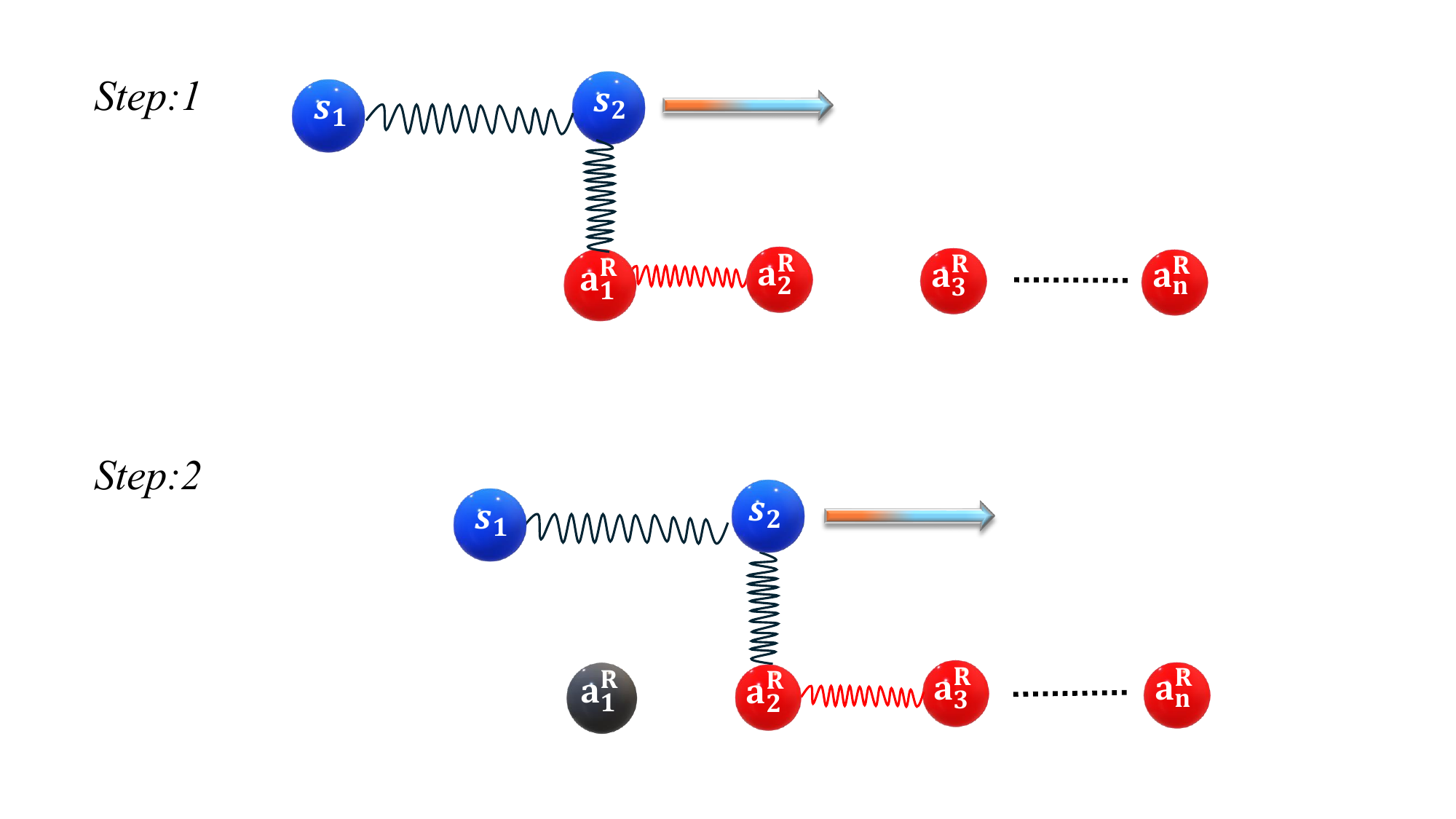}
    \caption{Schematic illustration of Scheme A, where only one system qubit $s_2$ interacts with a sequence of right-ancilla qubits $a_n^R$. In this diagram, the black lines represent Heisenberg-type interactions both between the two qubits of the system and between the system qubit $s_2$ and the ancillae. The red line indicates a partial-swap interaction occurring between successive ancilla qubits.}
    \label{approach_1}
\end{figure}

In this scheme, only the qubit $s_2$ of the two-qubit system $S$ interacts with the ancilla denoted by $a^R_n$ as shown in Fig.~\ref{approach_1}. The corresponding unitary concerning the intra-system and system-ancilla interaction is
\begin{equation}
    U_{s_1s_2, s_2a^R_n} = e^{-i H_A dt},
\end{equation}
where $H_A = H_{s_1} + H_{s_2} + H_{a_n^{R}} + H_{s_1s_2} + H_{s_2a^R_n}$ with $dt$ being the collision time step. Further, the intra-ancilla interaction unitary, $U_{a^R_na^R_{n+1}}$, is given by Eq.~\eqref{p_swap_unitary}. Now, the system's reduced state after the $n^{th}$ step of the scheme A is given by
\begin{equation}
    \rho_{n+1}^S = \Tr_{a^R_na^R_{n+1}}[\sigma^{Sa^R}_{n + 1}],
\end{equation}
where the joint state $\sigma_{S a_R^{n+1}}$ of the two-qubit system $S$ and the environmental ancillae $a_n^R$ after evolving through unitaries $U_{s_1s_2, s_2a^R_n}$ and $U_{a^R_na^R_{n+1}}$ is given by
\begin{equation}
    \sigma^{Sa^R}_{n + 1} = U_{a^R_na^R_{n+1}} U_{s_1s_2, s_2a^R_n}(\rho_n^{Sa_n^R} \otimes \rho^{a_{n + 1}^R})U_{s_1s_2, s_2a^R_n}^{\dag} U_{a^R_na^R_{n+1}}^{\dag}.
    \label{joint_state}
\end{equation}
For $n=1$, the initial joint state is $\rho^{Sa^R_1}_1 = \rho^S_0 \otimes \rho^{a^R_1}$. Additionally, $\rho_n^{Sa_n^R} = \Tr_{a^R_{n-1}}[\sigma^{Sa^R}_{n}]$ ensures that correlations are established between $S$ and $a_n^R$ on the $(n - 1)^{th}$ step, i.e., prior to their direct interaction on the $n^\text{th}$ collision.

\subsection{Scheme B}
\begin{figure}
    \centering
    \includegraphics[width=1\linewidth]{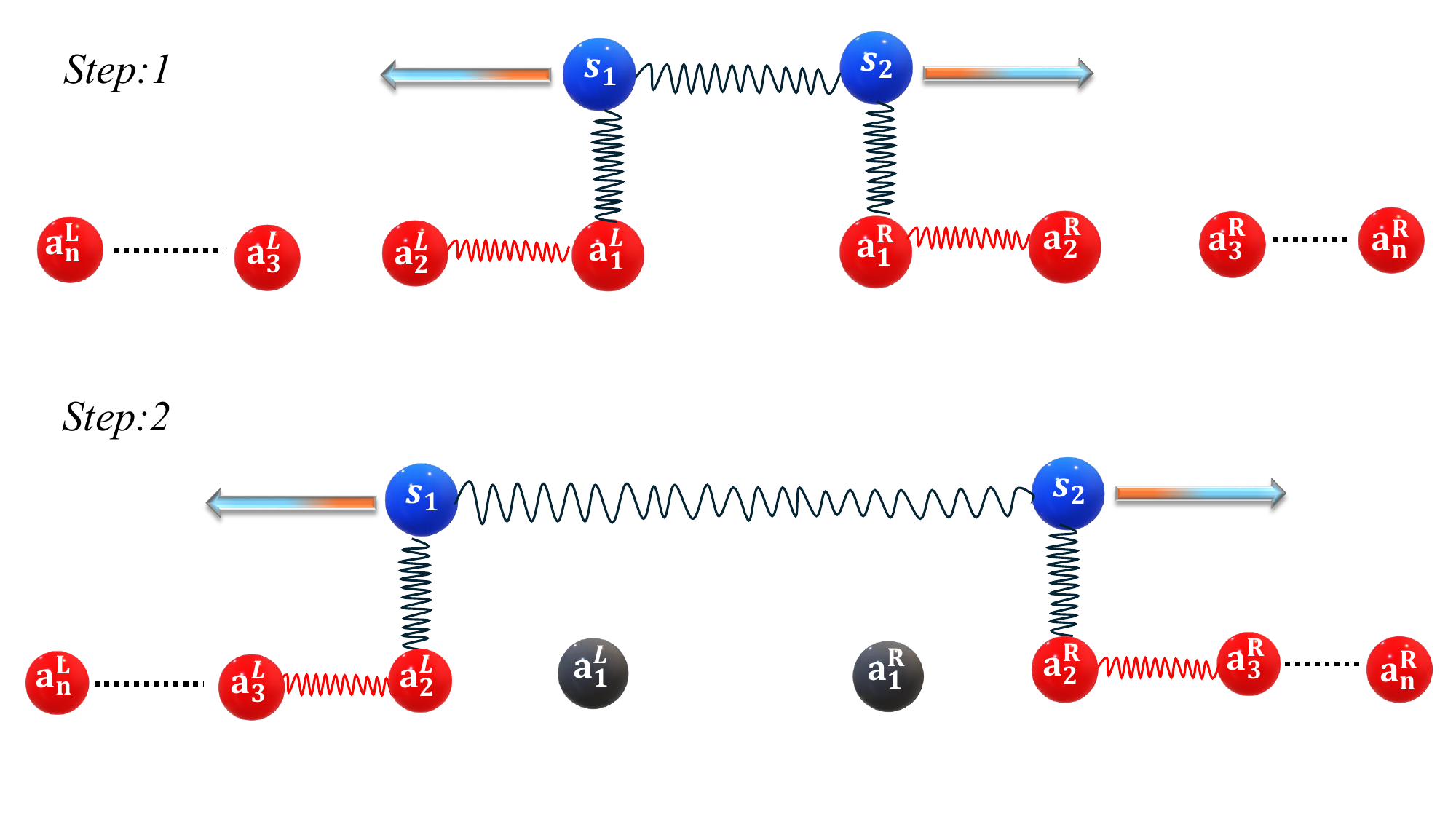}
    \caption{Schematic representation of Scheme B, where system qubits $s_1$ and $s_2$ interact with two independent streams of ancillae $a_n^L$ and $a_n^R$, respectively. In this diagram, black lines indicate Heisenberg-type interactions between the system qubits as well as between each system qubit and its corresponding ancilla. The red line represents a partial swap interaction occurring between successive ancillae in the same stream.}
    \label{Scheme_B}
\end{figure}

In this scheme, qubits $s_1$ and $s_2$ of the two-qubit system $S$ interact with the environment ancillae denoted by $a^L_n$ and $a^R_n$, respectively, as depicted in Fig.~\ref{Scheme_B}. The system's reduced state after the $n^{th}$ step of scheme B is 
\begin{equation}
    \rho_{n+1}^S = \Tr_{a^R_na^R_{n+1}a^L_na^L_{n+1}}[\sigma^{Sa^La^R}_{n + 1}],
\end{equation}
where
\begin{align}
    \sigma^{Sa^La^R}_{n + 1} = U_{a^R_na^R_{n+1}} U_{a^L_na^L_{n+1}} U_{s_1s_2, s_1a^L_n, s_2a^R_n}(\rho_n^{Sa_n^La_n^R} \otimes \rho^{a_{n + 1}^L} \nonumber \\
    \otimes \rho^{a_{n + 1}^R}) U_{s_1s_2, s_1a^L_n, s_2a^R_n}^{\dag} U_{a^L_na^L_{n+1}}^{\dag} U_{a^R_na^R_{n+1}}^{\dag},
\end{align}
and $\rho_n^{Sa_n^La_n^R} = \Tr_{a^R_{n-1}a^L_{n-1}}[\sigma^{Sa^La^R}_{n}]$.

Here, $\sigma^{sa^La^R}_{n+1}$ is the joint state of the two-qubit system and the environment ancillae after evolving through the unitary matrix corresponding to the intra-system and system environment interactions $U_{s_1s_2, s_1a^L_n, s_2a^R_n} = e^{-iH_Bdt}$, where $H_B = H_{s_1} + H_{s_2} + H_{a_n^{L}} + H_{a_n^{R}} + H_{s_1s_2} + H_{s_1a^L_n} + H_{s_2a^R_n}$, followed by the left side intra-ancilla interaction unitary $U_{a^L_na^L_{n+1}}$, and right side intra-ancilla interaction unitary $U_{a^R_na^R_{n+1}}$. The forms of $U_{a^L_na^L_{n+1}}$ and $U_{a^R_na^R_{n+1}}$ are as provided in Eq.~\eqref{p_swap_unitary}. For $n=1$, the initial joint state of the system and ancillae is $\rho^{Sa_1^La_1^R}_1 = \rho^S_0 \otimes \rho^{a^L_1} \otimes \rho^{a^R_1}$. 

\section{\label{Non_Markovianity_measure}Non-Markovianity witness}
\begin{figure}
    \centering
    \includegraphics[height=65mm,width=1\columnwidth]{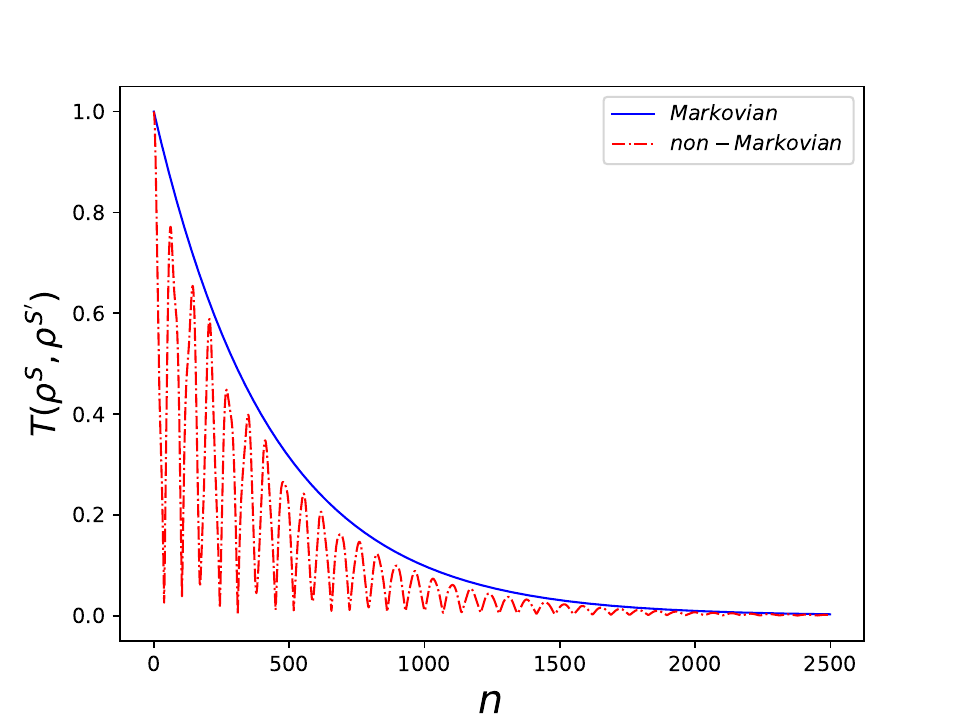}
    \caption{The evolution of the trace distance between the Bell states $\ket{\phi^{+}}$ and $\ket{\phi^{-}}$ is analyzed using the scheme A as a function of the number of collisions. In this analysis, the parameters are set as follows: $\omega_{s_1} = \omega_{s_2} = \omega_{a^R_n} = 1$, $g_{s_2a^R_n} = 0.85$, $g_{s_1s_2} = 0.95$, $dt = 0.08$ and $\beta_{a_n^{R}} = 1$. For Markovian dynamics, intra-ancilla interaction strength $\Theta = 0$, and for non-Markovian dynamics $\Theta = 0.95\pi/2$.}
    \label{Trace_distance_scheme_A}
\end{figure}

\begin{figure}
    \centering
    \includegraphics[height=40mm,width=1\columnwidth]{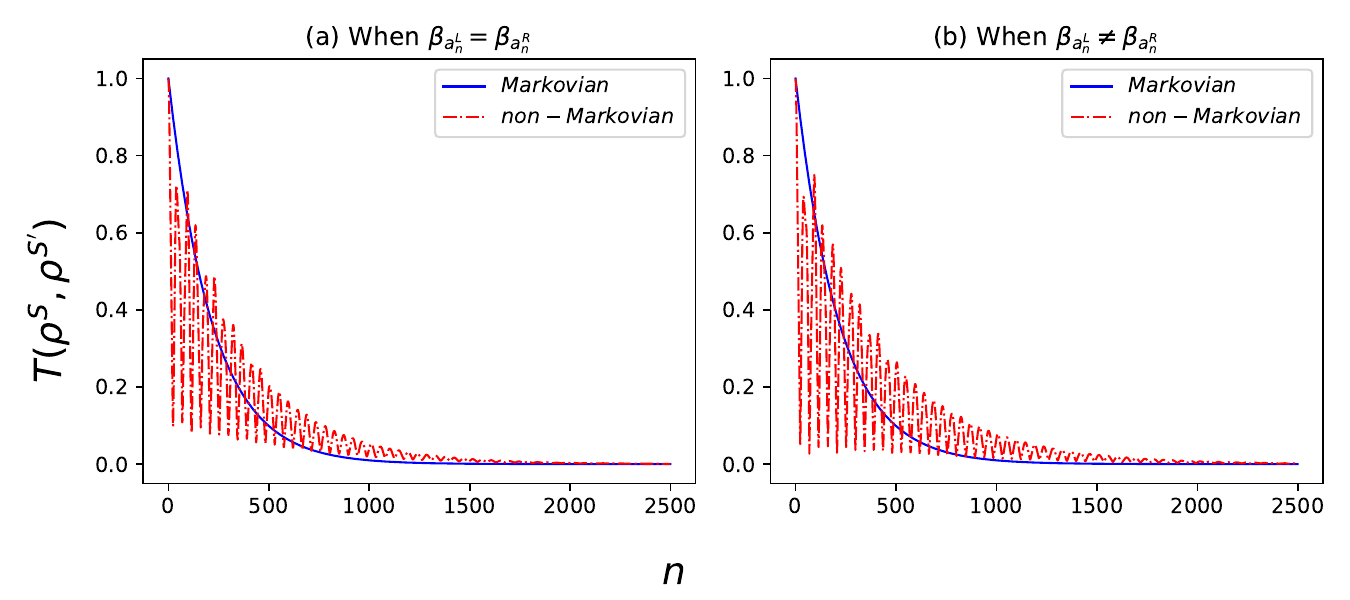}
    \caption{The evolution of the trace distance between the Bell states $\ket{\phi^{+}}$ and $\ket{\phi^{-}}$ is analyzed using the scheme B as a function of the number of collisions. In this analysis, the parameters are set as follows: $\omega_{s_1} = \omega_{s_2} = \omega_{a^L_n} = \omega_{a^R_n} = 1$, $g_{s_2a^R_n} = g_{s_1a^L_n} = 0.85$, $g_{s_1s_2} = 0.95$, $dt = 0.08$. In subplot (a) $\beta_{a_n^{L}} = \beta_{a_n^{R}} = 1$, and in subplot (b) $\beta_{a_n^{L}} = 1$, $\beta_{a_n^{R}} = 4$. For Markovian dynamics, intra-ancilla interaction strength $\Theta = 0$, and for non-Markovian dynamics $\Theta = 0.95\pi/2$.}
    \label{Trace_distance_scheme_B}
\end{figure}

The characterization and quantification of non-Markovianity in open quantum systems have garnered significant interest in recent times. Various measures for non-Markovianity have been introduced in literature, which claim to identify memory effects in the open systems' dynamics~\cite{rivas2014quantum, Breuer2016Colloquium, Utagi2020}. We study one of the extensively used non-Markovianity quantifiers, Breuer-Laine-Piilo (BLP) measure~\cite{Breuer2009Measure}. To identify memory effects that come from the non-Markovian nature of the open system dynamics, this method uses the distinguishability of the system's states based on their trace distance from one another. The trace distance $T$ is defined by
\begin{align}
    T(\rho^{S}_{n+1}, \rho^{S'}_{n+1}) &= \frac{1}{2}|| \rho^{S}_{n+1} - \rho^{S'}_{n+1} ||_1 \nonumber \\ 
    &= \frac{1}{2} \Tr\left[\sqrt{(\rho^{S}_{n+1} - \rho^{S'}_{n+1} )^{\dag} (\rho^{S}_{n+1} - \rho^{S'}_{n+1})}\right],
\end{align}
where $||.||_1$ represents the trace norm~\cite{nielsen2010quantum}, and $\rho^{S}_{n+1}$, $\rho^{S'}_{n+1}$ are the states of the two-qubit system at the $n^{th}$ step of scheme A and B, evolved from two different initial states of the system $S$ and $S'$, and depicted in Figs.~\ref{Trace_distance_scheme_A} and \ref{Trace_distance_scheme_B}, respectively. The flow of information between the open quantum system and its surroundings can be attributed to the change in the distinguishability between two arbitrary initial open quantum system states during the dynamics. 
If the distinguishability in the form of trace distance $T(\rho^{S}_{n+1}, \rho^{S'}_{n+1})$ experiences brief revivals over time evolution, it indicates an information backflow from the environment to the system, resulting in memory effects. Figures~\ref{Trace_distance_scheme_A} and~\ref{Trace_distance_scheme_B} show the variation of trace distance $T(\rho^{S}_{n+1}, \rho^{S'}_{n+1})$ with the number of collisions when the initial states of the system $S$ are Bell $\ket{\phi^{+}}$ and $\ket{\phi^{-}}$ states making use of schemes A and B, respectively. In all the cases, when the intra-ancilla interaction strength ($\Theta$) is zero, the trace distance monotonically decreases with time, depicting a Markovian evolution. For $\Theta = 0.95 \pi/2$, the trace distance shows an oscillatory behavior, depicting information backflow and a non-Markovian evolution.   

For the chosen parameters, the trace distance in scheme A remains consistently larger during the Markovian evolution than in the non-Markovian case, as can be seen from Fig.~\ref{Trace_distance_scheme_A}.   
In scheme B, we consider two cases. In the first case, both the environmental ancillae are at the same temperature, $\beta_{a^L_n} = \beta_{a^R_n}$, Fig.~\ref{Trace_distance_scheme_B}(a). In the second case, both are at different temperatures, $\beta_{a^L_n} \ne \beta_{a^R_n}$, Fig.~\ref{Trace_distance_scheme_B}(b). The initial state of the ancillae is according to Eq.~\eqref{eq_ancilla_thermal_state}. In the second case, where the temperatures are different, the oscillations in the trace distance are more prevalent, see Fig~\ref{Trace_distance_scheme_B}(b).

Having discussed the non-Markovian evolution of the two-qubit quantum collision model, we now study the non-classicality of the system using the Wigner function, non-classical volume, and entanglement generation. 

\section{\label{Wigner_function} Wigner function}
The Wigner function, $W(\theta, \phi)$, is a phase-space representation of a quantum state that extends classical probability distributions to the quantum domain~\cite{Wigner1932Quantum}. It is a real-valued function that can take negative values, indicating non-classicality~\cite{zavatta2004quantum}. A quasi-probability distribution can be described as a function of the polar $(\theta)$ and azimuthal $(\phi)$ angles using the relationship between spin-like, $SU(2)$ systems and the sphere. When this is extended over the entire basis set, with the spherical harmonics, the $W(\theta, \phi)$ function for a single spin-$j$ state can be written as~\cite{Thapliyal2015Quasiprobability}
\begin{equation}\label{Eq_Wigner1}
    W(\theta, \phi) = \sqrt{\frac{2j + 1}{4\pi}} \sum_{K, Q}\rho_{KQ}Y_{KQ}(\theta, \phi),
\end{equation}
where $K = 0, 1, ..., 2j$, and $Q = -K, -K+1, ....., 0, ....., K-1, K$,~$Y_{KQ}$ are the spherical harmonics and $\rho_{KQ} = \Tr\left[T_{KQ}^{\dag}\rho\right]$, with $\rho$ being the density matrix of the system. Further, multipole operators $T_{KQ}$ are given by
\begin{equation}
    T_{KQ} = \sum_{m, m'} (-1)^{j-m} (2K + 1)^{1/2} \begin{pmatrix}
                                j & K & j\\
                                -m & Q & m'
                               \end{pmatrix} 
                                \ket{j, m}\bra{j, m'},
\end{equation}
where $\begin{pmatrix} 
        j_1 & j_2 & j\\ 
        m_1 & m_2 & m 
        \end{pmatrix} = \frac{(-1)^{j_1 - j_2 - m}}{\sqrt{2j + 1}} \langle j_1m_1j_2m_2|j-m\rangle $ 
is the Wigner $3j$ symbol and $\langle j_1m_1j_2m_2|j-m\rangle$ is the Clebsh-Gordon coefficient. The multipole operators for $j = \frac{1}{2}$ are given in~\cite{Thapliyal2015Quasiprobability}. The Wigner function is also normalized such that $\int W(\theta, \phi) \sin{\theta} d{\theta} d{\phi} = 1$. Likewise, for a two-qubit system each with spin-$j$, the case considered here, the $W(\theta, \phi)$ function is 
\begin{align}
    W({\theta_1}, {\phi_1}, {\theta_2}, {\phi_2}) &= \left(\frac{2j + 1}{4\pi}\right)\sum_{K_1, Q_1}\sum_{K_2, Q_2}\rho^S_{n+1 {K_1}{Q_1}{K_2}{Q_2}}\nonumber \\
    &\times Y_{{K_1}{Q_1}}(\theta_1, \phi_1)Y_{K_2Q_2}{(\theta_2, \phi_2)},
    \label{two_qubit_W_func}
\end{align}
where $\rho^S_{n+1 K_1 Q_1 K_2 Q_2} = \Tr\left[\rho^S_{n+1} T_{K_1 Q_1}^{\dag} T_{K_2 Q_2}^{\dag}\right]$ and it is also normalized for $\theta_1, \theta_2 \in [0, \pi]$, and $\phi_1, \phi_2 \in [0, 2\pi]$.

\begin{figure}
    \centering
    \includegraphics[height=40mm,width=1\columnwidth]{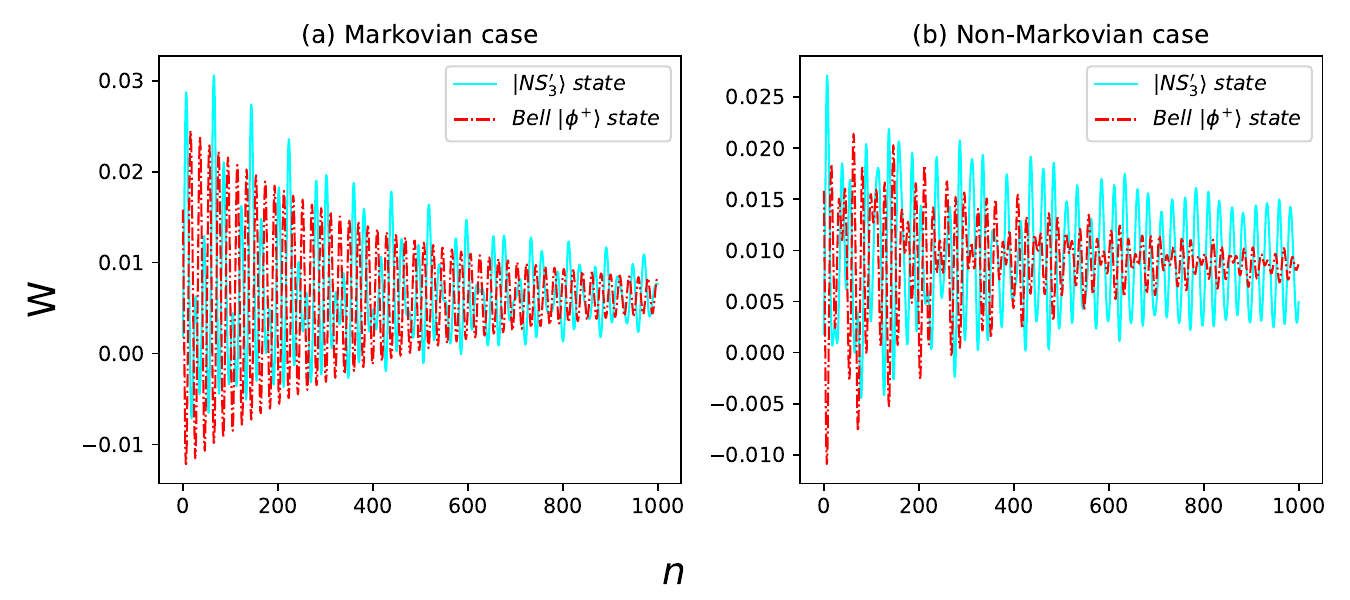}
    \caption{Variation of the Wigner function of the $\ket{NS_3^{\prime}}$, and Bell $\ket{\phi^{+}}$ states using the scheme A with the number of collisions. Here $\theta_1 = \theta_2 = \pi/2$, $\phi_1 = \phi_2 = \pi/6$, $\omega_{s_1} = \omega_{s_2} = \omega_{a^R_n} = 1$, $g_{s_2a^R_n} = 0.85$, $g_{s_1s_2} = 0.95$, $dt = 0.08$ and $\beta_{a_n^{R}} = 1$. Subplot (a) is for Markovian dynamics when $\Theta = 0$, and subplot (b) is for non-Markovian dynamics when $\Theta = 0.95 \times \pi/2$.}
    \label{W_func_H_int_two_qubit_aR_all_states_scheme_A}
\end{figure}

\begin{figure}
    \centering
    \includegraphics[height=68mm,width=1\columnwidth]{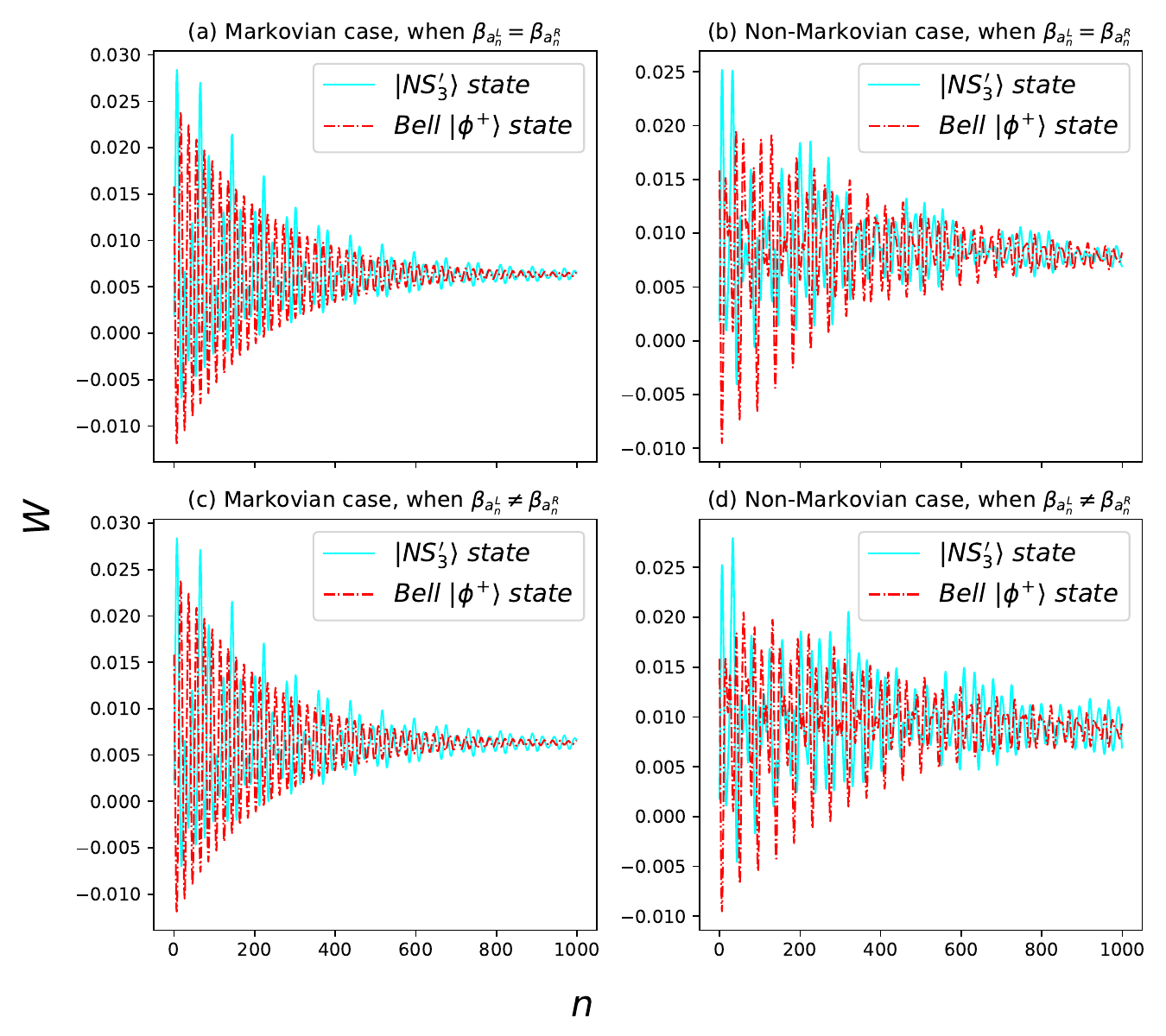}
    \caption{Variation of the Wigner function of the $\ket{NS_3^{\prime}}$, and Bell $\ket{\phi^{+}}$ states using the scheme B with the number of collisions. In this analysis, the parameters are set as follows: $\theta_1 = \theta_2 = \pi/2$, $\phi_1 = \phi_2 = \pi/6$, $\omega_{s_1} = \omega_{s_2} = \omega_{a^L_n} = \omega_{a^R_n} = 1$, $g_{s_2a^R_n} = g_{s_1a^L_n} = 0.85$, $g_{s_1s_2} = 0.95$, $dt = 0.08$. In subplot (a) and (b) $\beta_{a_n^{L}} = \beta_{a_n^{R}} = 1$, and in subplot (c) and (d) $\beta_{a_n^{L}} = 1$, $\beta_{a_n^{R}} = 4$. For Markovian dynamics, intra-ancilla interaction strength $\Theta = 0$, and for non-Markovian dynamics $\Theta = 0.95\pi/2$.}
    \label{W_func_H_int_two_qubit_aR_all_states_scheme_B}
\end{figure}

Figures~\ref{W_func_H_int_two_qubit_aR_all_states_scheme_A},~\ref{W_func_H_int_two_qubit_aR_all_states_scheme_B} illustrate the variation of Wigner function of the two-qubit $\ket{NS_3^{\prime}}$ state~\cite{lalita2023harnessing, Lalita_2024ProtectingQC, lalita2025realizingnegativequantumstates} (detailed in Appendix~\ref{NQS}) and Bell $\ket{\phi^{+}}$ state with the number of collisions using the schemes A and B, respectively. The $\ket{NS_3^{\prime}}$ is one of the two-qubit negative quantum states~\cite{lalita2023harnessing, Lalita_2024ProtectingQC, lalita2025realizingnegativequantumstates} known to offer enhanced robustness against non-Markovian noise for reliable quantum information processing tasks such as teleportation in realistic, noisy environments.

In scheme A, Fig.~\ref{W_func_H_int_two_qubit_aR_all_states_scheme_A}(b), for the non-Markovian evolution, the amplitude of oscillation of the Wigner function persists longer for the $\ket{NS_3^{\prime}}$ state in comparison to the Bell $\ket{\phi^{+}}$ state. Also, decay in amplitude of the Wigner function for the $\ket{NS_3^{\prime}}$ and $\ket{\phi^{+}}$ states in the non-Markovian case is comparatively lesser than the Markovian case as depicted by Fig. \ref{W_func_H_int_two_qubit_aR_all_states_scheme_A} (a) and (b). In Scheme B, Fig.~\ref{W_func_H_int_two_qubit_aR_all_states_scheme_B}, the Wigner function exhibits similar qualitative behavior for both identical and different ancilla temperatures. The decay of the amplitude of oscillation for the Wigner function is gradual and relatively ordered in the Markovian case than in the non-Markovian scenario, see Fig. \ref{W_func_H_int_two_qubit_aR_all_states_scheme_B} (a), (b), (c), and (d).  

In all the cases, we observe that the variation of the Wigner function with the number of collisions for both the two-qubit states shows negative values, indicating quantumness. Now, to quantify the quantumness, we study the non-classical volume.  

\subsection{Non-classical Volume}
The Wigner function's negative values serve as a non-classicality signature. Nevertheless, the negative values do not offer a quantitative measure of non-classicality. The non-classical volume, introduced in~\cite{Anatole2004Negativity}, is used as a quantitative measure of quantumness in a given quantum system. For a two-qubit quantum system, it is given by
\begin{equation}
    \delta = \int |W({\theta_1}, {\phi_1}, {\theta_2}, {\phi_2})| \sin({\theta_1}) \sin({\theta_2}) d{\theta_1} d{\theta_2} d{\phi_1} d{\phi_2} - 1,
\end{equation}
where $W({\theta_1}, {\phi_1}, {\theta_2}, {\phi_2})$ is the Wigner function for a two-qubit quantum system defined above in Eq.~\eqref{two_qubit_W_func}. A non-zero value of $\delta$ indicates that the system is non-classical. 

\begin{figure}
    \centering
    \includegraphics[height=40mm,width=1\columnwidth]{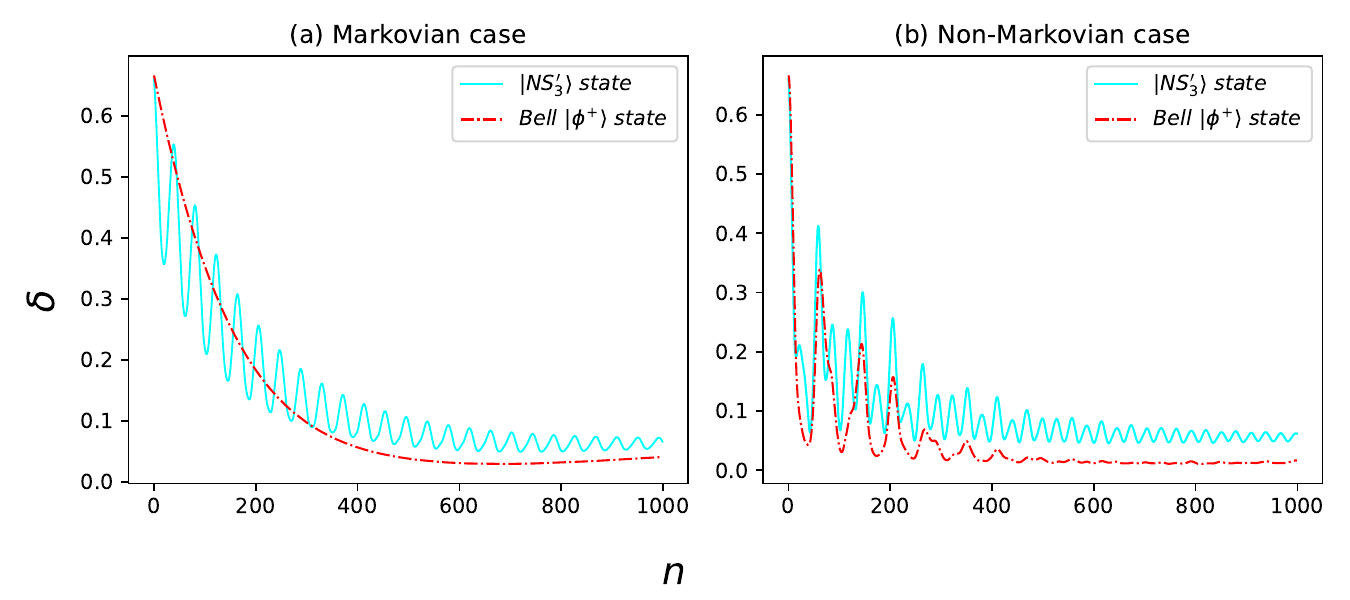}
    \caption{Variation of the non-classical volume of the $\ket{NS_3^{\prime}}$, and Bell $\ket{\phi^{+}}$ states with the number of collisions using scheme A. Here $\omega_{s_1} = \omega_{s_2} = \omega_{a^R_n} = 1$, $g_{s_2a^R_n} = 0.85$, $g_{s_1s_2} = 0.95$, $dt = 0.08$ and $\beta_{a_n^{R}} = 1$. Subplot (a) is for Markovian dynamics when $\Theta = 0$, and subplot (b) is for non-Markovian dynamics when $\Theta = 0.95 \times \pi/2$.}
    \label{non_classical_vol_H_int_two_qubit_aR_all_states_scheme_A}
\end{figure}

\begin{figure}
    \centering
    \includegraphics[height=65mm,width=1\columnwidth]{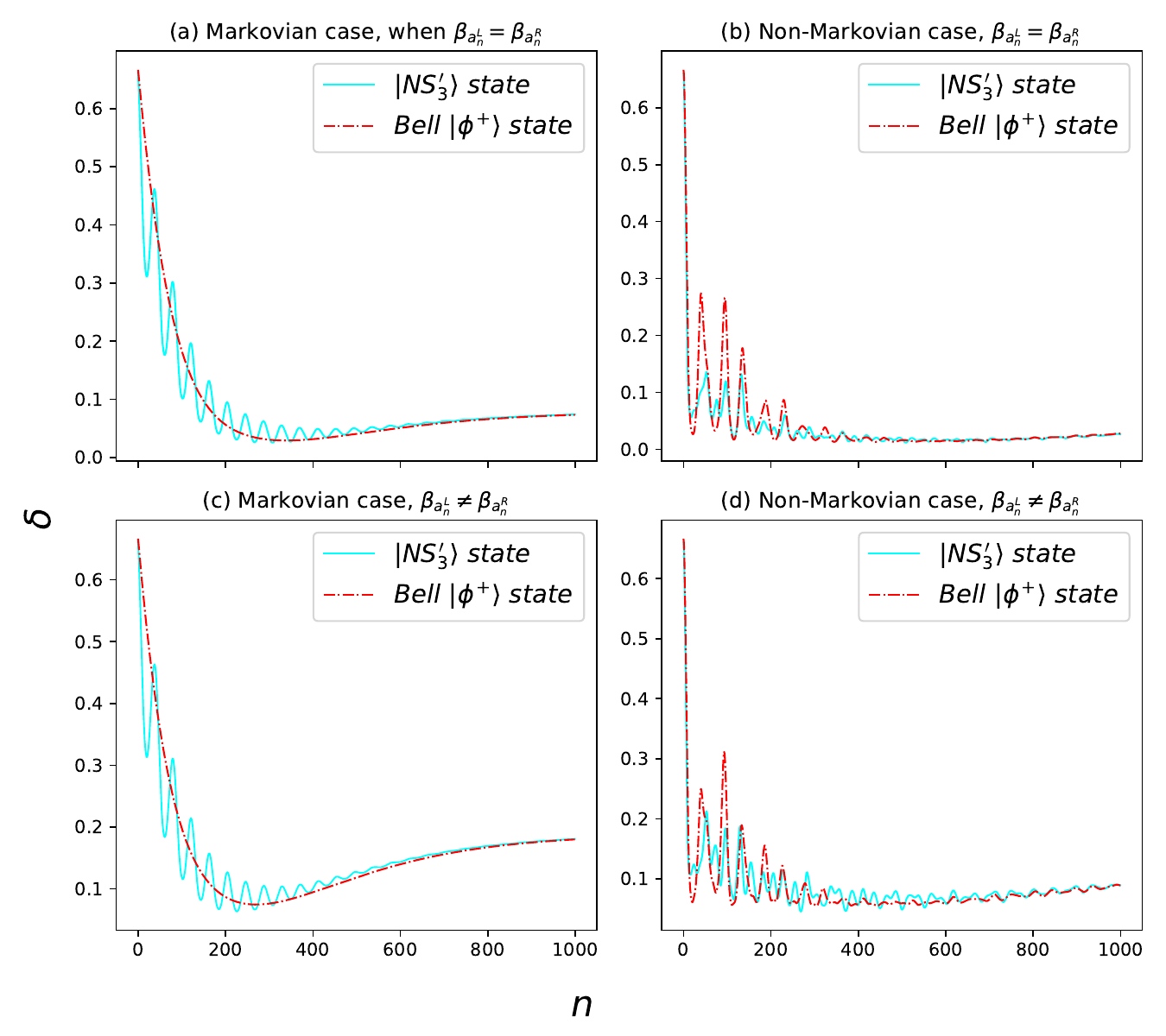}
    \caption{Variation of the non-classical volume of the $\ket{NS_3^{\prime}}$, and Bell $\ket{\phi^{+}}$ states with the number of collisions using scheme B. In this analysis, the parameters are set as follows: $\omega_{s_1} = \omega_{s_2} = \omega_{a^L_n} = \omega_{a^R_n} = 1$, $g_{s_2a^R_n} = g_{s_1a^L_n} = 0.85$, $g_{s_1s_2} = 0.95$, $dt = 0.08$. In subplot (a) and (b) $\beta_{a_n^{L}} = \beta_{a_n^{R}} = 1$, and in subplot (c) and (d) $\beta_{a_n^{L}} = 1$, $\beta_{a_n^{R}} = 4$. For Markovian dynamics, intra-ancilla interaction strength $\Theta = 0$, and for non-Markovian dynamics $\Theta = 0.95\pi/2$.}
    \label{non_classical_vol_H_int_two_qubit_aR_all_states_scheme_B}
\end{figure}

The variation of non-classical volume of the $\ket{NS_3^{\prime}}$ and Bell $\ket{\phi^{+}}$ states, using the scheme A and B, is depicted in Figs.~\ref{non_classical_vol_H_int_two_qubit_aR_all_states_scheme_A},~\ref{non_classical_vol_H_int_two_qubit_aR_all_states_scheme_B}, respectively, for both (non-)Markovian evolution. 
We observe that for all the cases, both states have the same initial value of non-classical volume. The variation of non-classical volume for the Bell state is oscillatory in the non-Markovian scenario and non-oscillatory in the Markovian scenario of both schemes A and B.
Interestingly, the $\ket{NS_3^{\prime}}$ state's non-classical volume shows oscillatory behavior with the number of collisions, even in the Markovian case, see Figs.~\ref{non_classical_vol_H_int_two_qubit_aR_all_states_scheme_A}(a),~\ref{non_classical_vol_H_int_two_qubit_aR_all_states_scheme_B}(a) and (c). This happens because the state (see Appendix~\ref{NQS}) has complex numbers in it. In scheme A of the collision model, the $\ket{NS_3^{\prime}}$ state sustains its non-classical behavior longer in both the Markovian and non-Markovian cases, as can be seen from Fig.~\ref{non_classical_vol_H_int_two_qubit_aR_all_states_scheme_A}. On the other hand, in scheme B, at longer duration, both states show equivalent variation of non-classical volume, Fig.~\ref{non_classical_vol_H_int_two_qubit_aR_all_states_scheme_B}.

We now study how schemes A and B and (non-)Markovian evolution affect quantum correlations in the system qubits.  

\section{\label{Quantum_correlation} Quantum correlations}
Quantum non-local correlations are one of the most profound and intrinsically non-classical features of a quantum system, exhibiting phenomena that lack any counterpart in classical physics. Central to these correlations is entanglement, which serves as a foundational element in describing quantum correlations within composite quantum systems. Specifically, in the case of a bipartite system composed of two qubits, entanglement can be quantitatively characterized by concurrence \cite{Wootters1998Entanglement}
\begin{equation}
   \begin{aligned}
     C({\rho}^S_{n+1}) = \max \{0, \lambda_{1} - \lambda_{2} - \lambda_{3} - \lambda_{4}\},
      \end{aligned}
      \label{concur_eq.}
\end{equation}
where $\lambda_{i}$'s are the eigenvalues of $\sqrt{\sqrt{{\rho}^S_{n+1}} \tilde{{\rho}}^S_{n+1} \sqrt{{\rho}^S_{n+1}}}$, such that $\lambda_{1} \geq \lambda_{2} \geq \lambda_{3} \geq \lambda_{4}$, and $\Tilde{{\rho}}^S_{n+1} = (\sigma_{y} \otimes \sigma_{y}) {\rho}_{n+1}^{S*} (\sigma_{y} \otimes \sigma_{y})$, where ${\rho}_{n+1}^{S*}$ is the complex conjugate of ${\rho}^S_{n+1}$. Further, to investigate whether entanglement can emerge between initially separable qubits of the system as a result of sufficiently strong intra-system and system-ancilla interaction strengths, we analyze how the concurrence evolves with the number of collisions of the two-qubit collision model. 

\begin{figure}
    \centering
    \includegraphics[height=40mm,width=1\columnwidth]{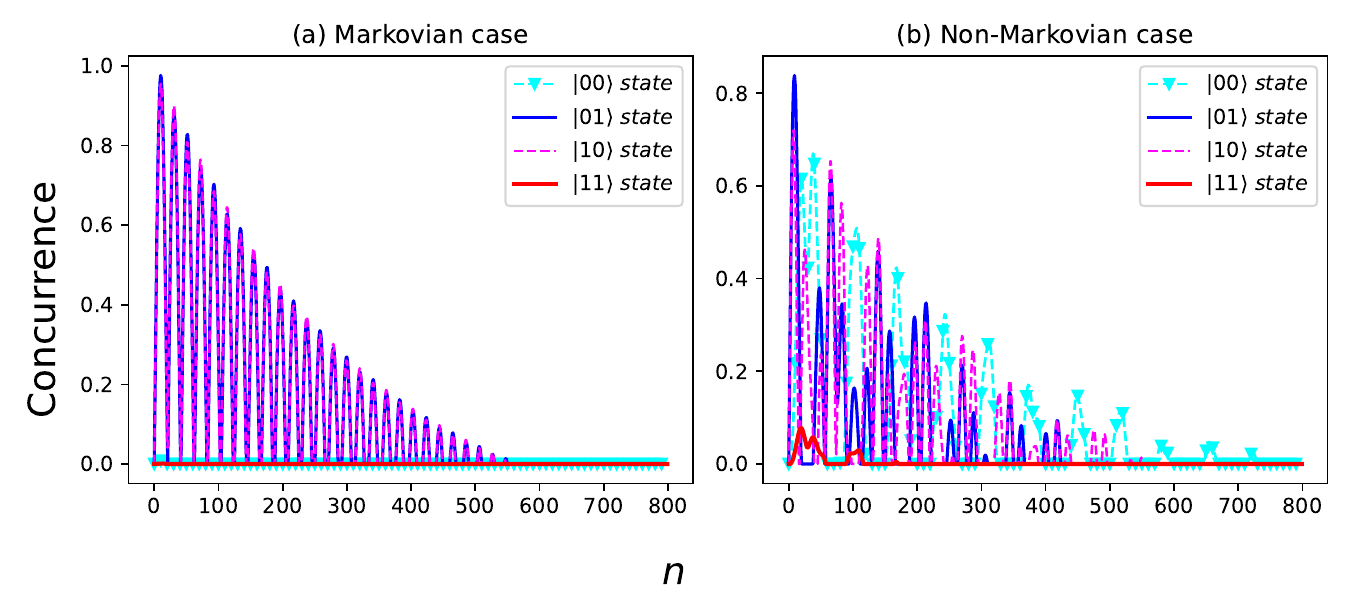}
    \caption{Variation of the concurrence of the $\ket{00}$, $\ket{01}$, $\ket{10}$ and $\ket{11}$ states using the scheme A with the number of collisions. Here $\omega_{s_1} = \omega_{s_2} = \omega_{a^R_n} = 1$, $g_{s_2a^R_n} = 0.85$, $g_{s_1s_2} = 0.95$, $dt = 0.08$ and $\beta_{a_n^{R}} = 1$. Subplot (a) is for Markovian dynamics when $\Theta = 0$, and subplot (b) is for non-Markovian dynamics when $\Theta = 0.95 \times \pi/2$.}
    \label{concurrence_H_int_two_qubit_aR_all_states_scheme_A}
\end{figure}

\begin{figure}
    \centering
    \includegraphics[height=65mm,width=1\columnwidth]{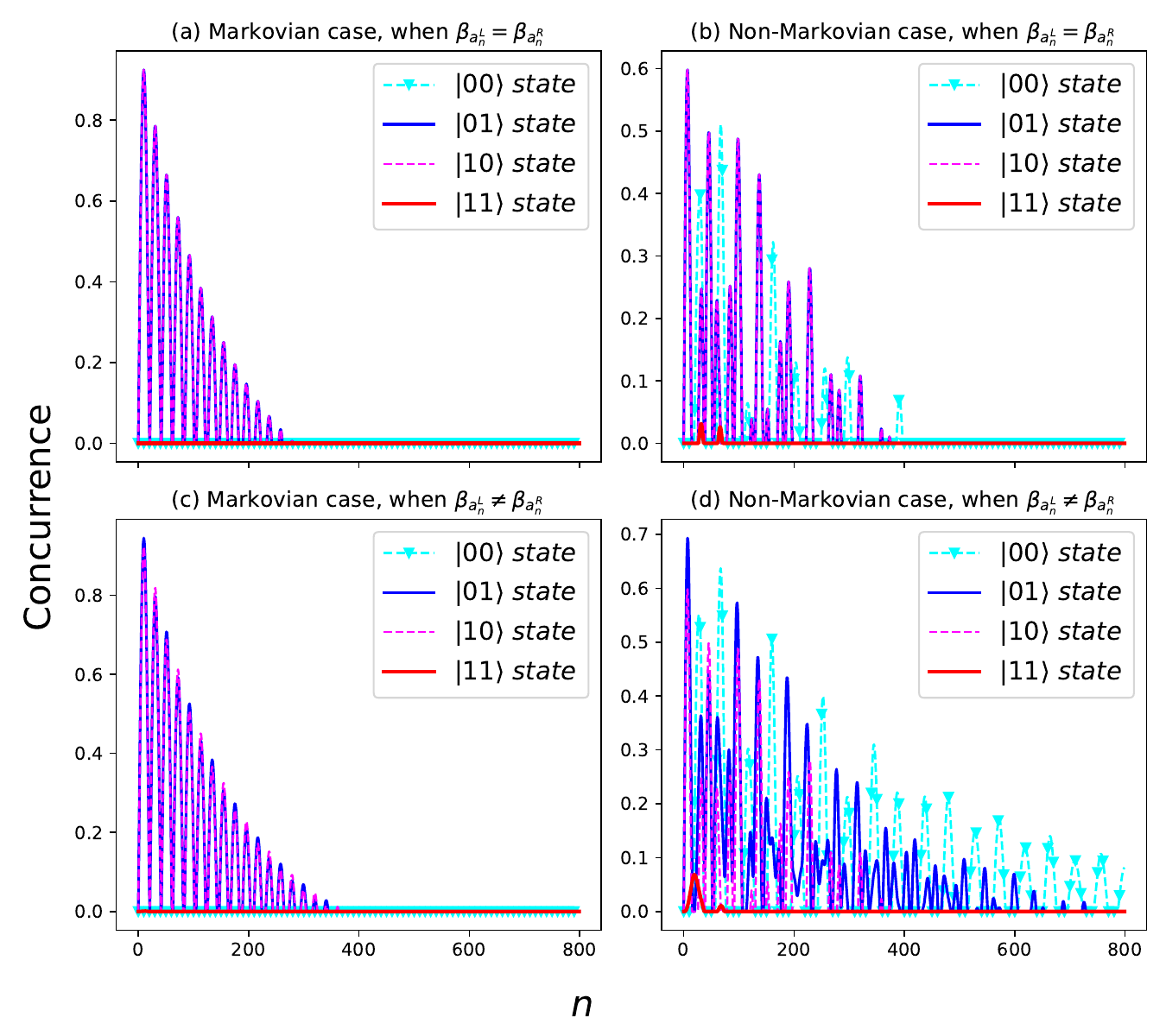}
    \caption{Variation of the concurrence of the $\ket{00}$, $\ket{01}$, $\ket{10}$ and $\ket{11}$ states with the number of collisions using scheme B. In this analysis, the parameters are set as follows: $\omega_{s_1} = \omega_{s_2} = \omega_{a^L_n} = \omega_{a^R_n} = 1$, $g_{s_2a^R_n} = g_{s_1a^L_n} = 0.85$, $g_{s_1s_2} = 0.95$, $dt = 0.08$. In subplot (a) and (b) $\beta_{a_n^{L}} = \beta_{a_n^{R}} = 1$, and in subplot (c) and (d) $\beta_{a_n^{L}} = 1$, $\beta_{a_n^{R}} = 4$. For Markovian dynamics, intra-ancilla interaction strength $\Theta = 0$, and for non-Markovian dynamics $\Theta = 0.95\pi/2$.}
    \label{concurrence_H_int_two_qubit_aR_aL_all_states_scheme_B}
\end{figure}

Figures~\ref{concurrence_H_int_two_qubit_aR_all_states_scheme_A} and~\ref{concurrence_H_int_two_qubit_aR_aL_all_states_scheme_B} illustrate the variation of concurrence with the number of collisions using scheme A and B, respectively, for the $\ket{00}$, $\ket{01}$, $\ket{10}$ and $\ket{11}$ states as the initial separable states of the two-qubit system. In the Markovian regime, for both schemes A and B, no appreciable entanglement is generated for the $\ket{00}$ and $\ket{11}$ states under the specified intra-system and system-ancilla interaction strengths, see Figs.~\ref{concurrence_H_int_two_qubit_aR_all_states_scheme_A}(a), ~\ref{concurrence_H_int_two_qubit_aR_aL_all_states_scheme_B}(a), and~\ref{concurrence_H_int_two_qubit_aR_aL_all_states_scheme_B}(c). Interestingly, in the non-Markovian regime, for both schemes A and B, there is significant entanglement generation for the $\ket{00}$ state, as can be seen from Figs.~\ref{concurrence_H_int_two_qubit_aR_all_states_scheme_A}(b),~\ref{concurrence_H_int_two_qubit_aR_aL_all_states_scheme_B}(b), and  ~\ref{concurrence_H_int_two_qubit_aR_aL_all_states_scheme_B}(d). Also, in the non-Markovian regime, for $\ket{11}$ initial state, entanglement is seen to be generated initially, but it is not substantial.

However, substantial entanglement growth, as quantified by concurrence, is observed for the $\ket{01}$ and $\ket{10}$ states in both (non-)Markovian regimes of schemes A and B. This sustains for a few hundred collisions before it becomes zero, because of open system effects.  Although the $\ket{01}$ and $\ket{10}$ states initially display the highest concurrence, the $\ket{00}$ state shows more sustained entanglement over time in both schemes within the non-Markovian framework, see Figs.~\ref{concurrence_H_int_two_qubit_aR_all_states_scheme_A}(b), ~\ref{concurrence_H_int_two_qubit_aR_aL_all_states_scheme_B}(b), and~\ref{concurrence_H_int_two_qubit_aR_aL_all_states_scheme_B}(d). Furthermore, for the specified intra-ancilla and system-ancilla coupling strengths, scheme B yields a more robust and sustained entanglement over a larger number of collisions when the ancillae are maintained at different temperatures in the non-Markovian regime, as illustrated in Fig.~\ref{concurrence_H_int_two_qubit_aR_aL_all_states_scheme_B}(d).

In the subsequent section, we study the steady state behavior of the system qubits and the impact of ancilla temperature on both qubits of the two-qubit collision model.

\section{\label{SS_behaviour} Steady state behaviour}
The steady state of an open quantum system refers to the asymptotic state in which the system’s density matrix does not change due to repeated interactions with its environment~\cite{Breuer2007, Strasberg2017Quantum}. This state, often independent of the initial conditions, may correspond to thermal equilibrium (Gibbs state) or a non-equilibrium stationary state, depending on the system-environment dynamics. Observing whether the system relaxes to a Gibbs state of the form $\rho\propto e^{-\beta H_s}$ helps to evaluate the applicability of thermodynamic principles to small quantum systems. Here, we determine the steady state of a two-qubit system by evolving it under scheme B. The steady state is where the difference between successive reduced system states, $\rho^S_{n+1} - \rho^S_n$, vanishes; here $n$ denotes the $n^{th}$ collision. We then obtain the reduced density matrix of each qubit at this convergence point, which is referred to as the individual system's qubit steady state. Here, we keep the temperature of both ancillae to be the same $\beta_{a^L_n} = \beta_{a^R_n} = \beta$ and determine whether the individual qubit steady state corresponds to a thermal state of its bare system Hamiltonian. To this end, we calculate the fidelity between the steady state and the Gibbs states of each qubit, using the following formula~\cite{jozsa1994fidelity, nielsen2010quantum},
\begin{equation}
    F(\rho^{ss}_{s_{k}}, \sigma_{Gs_k}) \equiv {\rm Tr}\sqrt{\sqrt{\rho^{ss}_{s_{k}}}\sigma_{Gs_k}\sqrt{\rho^{ss}_{s_{k}}}},~~\forall~ k \in \{1, 2\},
    \label{Fidelity_formula}
\end{equation}
where $\rho^{ss}_{s_{k}}$ denotes the steady state of the $k$-th qubit of the system $S$ and $\sigma_{Gs_k} = e^{-\beta H_{s_{k}}}/\Tr[e^{-\beta H_{s_{k}}}]$ denotes the Gibbs states corresposing to their bare system Hamiltionians.

\begin{figure}
    \centering
    \includegraphics[height=45mm,width=1\columnwidth]{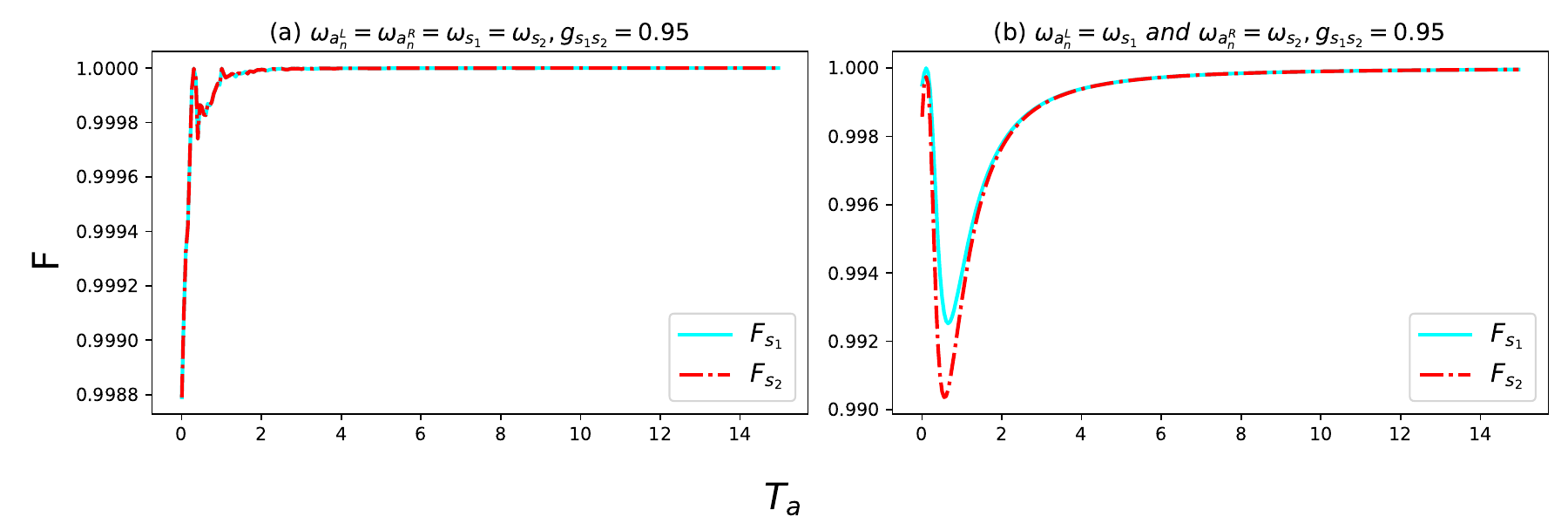}
    \caption{Evolution of fidelity between the each system qubit's steady state and the Gibbs state corresponding to their bare Hamiltonian with the temperature of ancillae, i.e., $T_{a^{L}_n} = 1/\beta_{a^{L}_n}$, and $T_{a^{R}_n} = 1/\beta_{a^{R}_n}$, under scheme B within the context of non-Markovian dynamics ($\Theta = 0.95\pi/2$). Here $T_{a^{L}_n} = T_{a^{R}_n} = T_a$. The parameters are set as follows: in subplot (a) $\omega_{s_1} = \omega_{s_2} = \omega_{a^L} = \omega_{a^R} = 1$,  $g_{s_1s_2} = 0.95$, in subplot (b) $\omega_{s_1} = \omega_{a^L} = 0.5$, $\omega_{s_2} = \omega_{a^R} = 1$, $g_{s_1s_2} = 0.95$ and for all plots $g_{s_1a^L} = g_{s_2a^R} = 0.5$, and $dt = 0.1$.}
    \label{ss_fidelity_all_cases}
\end{figure}

In Fig.~\ref{ss_fidelity_all_cases}, we plot the variation of fidelity with the temperature of the ancillae, when the initial state of the two-qubit system is $\ket{00}$. Fig.~\ref{ss_fidelity_all_cases}(a) corresponds to the case when all the ancillae and system qubits' transition frequencies are the same, while in Fig.~\ref{ss_fidelity_all_cases}(b), the transition frequencies of the qubits are different, however, the transition frequencies $\omega_{a^L_n} = \omega_{s_1}$ and $\omega_{a^R_n} = \omega_{s_2}$. In both Figs.~\ref{ss_fidelity_all_cases}(a) and (b), the coupling between the system qubits is taken to be of the order of their transition frequencies, i.e, around 0.95, denoting a strong coupling between them. Here, in the low temperature regime, we observe that the fidelity between the steady state and the Gibbs state corresponding to the bare system Hamiltonian is lower than one (where $F=1$ denotes that both the states are the same) and fluctuates. As the temperature is increased, the fidelity stabilizes and saturates at one. This denotes that the steady state becomes equal to the Gibbs state corresponding to the bare system Hamiltonian at higher temperatures. Usually, the thermal steady state of the system is given by the Gibbs state corresponding to an effective Hamiltonian, different from the bare system Hamiltonian at low temperatures and in the strong coupling regime~\cite{Hanggi_talkner_review}. This becomes equal to the bare system Hamiltonian as the coupling decreases and with higher temperature~\cite{Hanggi_talkner_review, Miller2018, tiwari2024strong, pathania2024}. 

\section{\label{conclusion}Conclusions}
The quantum collision model has emerged as a robust and conceptually simple framework for simulating open quantum dynamics of single-qubit systems. However, its extension to multipartite systems remains comparatively underexplored. In this work, we have explored a two-qubit quantum collision model. Two schemes were introduced to model the two-qubit collision model. In scheme A, a stream of ancillae interacted with only one of the qubits of the two-qubit system. In scheme B, both the qubits interacted with independent streams of ancillae. Here, an ancilla interacts with the system and subsequently interacts with the next ancilla destined to interact with the system. This interaction between ancillae was shown to introduce memory effects into the system. This non-Markovian feature was explored using the BLP measure, where the trace distance between two different initial states showed oscillatory evolution with the number of collisions. The non-classical features of the model were studied first using the Wigner function and non-classical volume. For this, the two-qubit negative quantum state $\ket{NS_3^\prime}$ and the Bell states were used as the initial states of the system. It was found that, in scheme A, the $\ket{NS_3^\prime}$ state was advantageous, as it was robust against the decay of non-classicality for a longer duration. Next, the behavior of quantum correlations, particularly entanglement, was studied using concurrence. In the Markovian case, entanglement was not significantly generated for the most excited and the ground states of the system. However, in the non-Markovian case, entanglement was generated for the most excited state, and it was robust for a larger number of collisions compared to the other states.

Furthermore, the system qubits' steady-state behavior was also analyzed. It was probed whether the thermal steady state coincides with the Gibbs state corresponding to the bare system Hamiltonian. This was found to be true in the high-temperature regime. However, in the low-temperature and strong coupling regime, this was not so, indicating that here the Gibbs state corresponding to the bare system's Hamiltonian is no longer equivalent to the thermal steady state of the system. This paper provides a comprehensive study of the two-qubit quantum collision model and opens up research avenues, such as thermalization in quantum thermodynamics in the multi-qubit system scenario.

\section*{Acknowledgments}
J. L. acknowledges Devvrat Tiwari and Baibhab Bose for their useful discussions and valuable insights.

\bibliography{BibTexfile}
\bibliographystyle{apsrev4-2}

\clearpage
\onecolumngrid
\appendix

\section{\label{NQS}{Negative Quantum states}}
Negative quantum states are defined as the normalized eigenvectors associated with the negative eigenvalues of the phase space point operator $A_\alpha$ at a given phase space coordinate $\alpha(q, p)$~\cite{lalita2023harnessing, Lalita_2024ProtectingQC, lalita2025realizingnegativequantumstates}. The operator $A_\alpha$ plays a pivotal role in discrete phase space formulations, as it directly determines the discrete Wigner function (DWF) via the relation $W_\alpha = \frac{1}{d} \text{Tr}(\rho A_\alpha)$, where $d$ is the system's Hilbert space dimension~\cite{van2011noise, casaccino2008extrema, lalita2023harnessing}. Particularly, the eigenvectors corresponding to the negative eigenvalues of $A_\alpha$ are known to minimize the DWF, thereby revealing signatures of quantum non-classicality~\cite{van2011noise, lalita2023harnessing}. Using the framework developed in~\cite{wootters2004picturing, gibbons2004discrete}, the negative quantum states are obtained for $d = 2, 3, 4$~\cite{lalita2023harnessing, Lalita_2024ProtectingQC}. Moreover, the eigenvector corresponding to the most negative eigenvalue of the phase space point operator $A_\alpha$ is referred to as the first negative quantum state and is denoted by $|NS_1\rangle$. Similarly, the second and third negative quantum states, denoted by $|NS_2\rangle$ and $|NS_3\rangle$, respectively, correspond to the normalized eigenvectors associated with the second and third most negative eigenvalues of $A_\alpha$. This pattern extends to further negative eigenvalues, yielding additional negative quantum states. 

In two-qubit systems, quantum states that show negativity in their discrete Wigner functions are highly robust against various types of noise, primarily those described by non-Markovian dynamics~\cite{lalita2023harnessing, Lalita_2024ProtectingQC}. These negative quantum states are resilient and ideal candidates for universal quantum teleportation using weak measurements in non-Markovian environments~\cite{Lalita_2024ProtectingQC}. To support their practical implementation, quantum circuits for generating these states have been designed and demonstrated using the Qiskit framework~\cite{lalita2025realizingnegativequantumstates, qiskit2024}. These circuits offer a feasible approach to realizing negative states as reliable resources in quantum information tasks. Approximate forms of several two-qubit negative quantum states are~\cite{lalita2023harnessing, Lalita_2024ProtectingQC, lalita2025realizingnegativequantumstates}.,
\[
\begin{aligned}
|NS_1\rangle &= \begin{pmatrix} a \\ b \\ c \\ d \end{pmatrix}, \quad
|NS_2\rangle = \begin{pmatrix} p \\ q \\ r \\ s \end{pmatrix}, \quad
|NS_3\rangle = \begin{pmatrix} -l \\ m \\ n \\ l \end{pmatrix}, \\
|NS_3'\rangle &= \begin{pmatrix} -x \\ y \\ z \\ x \end{pmatrix}, \quad
|NS_3''\rangle = \begin{pmatrix} 0 \\ ik \\ k \\ 0 \end{pmatrix},
\end{aligned}
\]
where the numerically rounded coefficients are:
\[
\begin{aligned}
&a = -0.743, \quad b = -0.357(1 - i), \quad c = 0.102(1 + i), \quad d = -0.414, \\
&p = 0.788, \quad q = -0.288(1 - i), \quad r = -0.288(1 + i), \quad s = -0.211, \\
&l = 0.0508, \quad m = 0.631 - 0.228i, \quad n = -0.279 - 0.682i, \\
&x = 0.575, \quad y = -0.346 + 0.310i, \quad z = -0.265 - 0.229i, ~\text{and}, \quad k = \frac{1}{\sqrt{2}}.
\end{aligned}
\]
These expressions represent numerically derived approximations of the two-qubit negative quantum states associated with distinct spectral profiles of the phase space point operators.

\end{document}